\documentclass[fleqn]{article}
\usepackage[a4paper,margin=25.4mm]{geometry}
\usepackage[english]{babel}

\usepackage{amsmath}
\usepackage{amssymb}
\usepackage{sistyle}
\usepackage{cite}

\usepackage{graphicx}			
\usepackage{wrapfig}			
\usepackage[
	labelfont={sf,bf},			
	textfont={sf},			
	font=small,					
	justification=RaggedRight	
	]{caption}
\usepackage{subcaption}			
\usepackage{epstopdf}			
\usepackage{epsfig}
\usepackage{url}

\newcommand{\mbf}[1]{\mathbf{#1}}

\begin{document}

\begin{center}
	\hfill\\[1.0cm]
	{\Large\textbf{Measurement of the transmission secondary electron yield of nanometer-thick films in a prototype Timed Photon Counter}}\\[0.2cm]
	{\large{T.H.A. van der Reep$\left.^{1,2}\right.$, B. Looman$\left.^{2}\right.$, H.W. Chan$\left.^{1,3}\right.$, C.W. Hagen$\left.^{2}\right.$ and\\ H. van der Graaf$\left.^{1,2,*}\right.$}} \\[0.2cm]
	{$\left.^{1}\right.$\emph{Nederlands Instituut voor Kern- en Hoge Energie Fysica (Nikhef), Science Park $105$, $1098$ XG Amsterdam, The Netherlands}}\\[0.1cm]
	{$\left.^{2}\right.$\emph{Faculty of applied sciences, Department of Imaging Physics, Delft University of Technology, Lorentzweg $1$, $2628$ CJ Delft, The Netherlands}}\\[0.1cm]
	{$\left.^{3}\right.$\emph{Faculty of electrical engineering, mathematics and computer science, Department of Microelectronics/ECTM, Delft University of Technology, Feldmanweg $17$, $2628$ CT Delft, Netherlands}}\\[0.1cm]
	$\left.^{*}\right.$vdgraaf@nikhef.nl\\[0.2cm]
	\today
\end{center}

\begin{abstract}
We measure the transmission secondary electron yield of nanometer-thick Al$_2$O$_3$/TiN/Al$_2$O$_3$ films using a prototype version of a Timed Photon Counter (TiPC). We discuss the method to measure the yield extensively. The yield is then measured as a function of landing energy between $1.2$ and $\SI{1.8}{keV}$ and found to be in the range of $0.1$ ($\SI{1.2}{keV}$) to $0.9$ ($\SI{1.8}{keV}$). These results are in agreement to data obtained by a different, independent method. We therefore conclude that the prototype TiPC is able to characterise the thin films in terms of transmission secondary electron yield. Additionally, observed features which are unrelated to the yield determination are interpreted.
\end{abstract}

\section{Introduction}
\begin{figure}[htbp]
	\centering
	\includegraphics[width=0.45\textwidth]{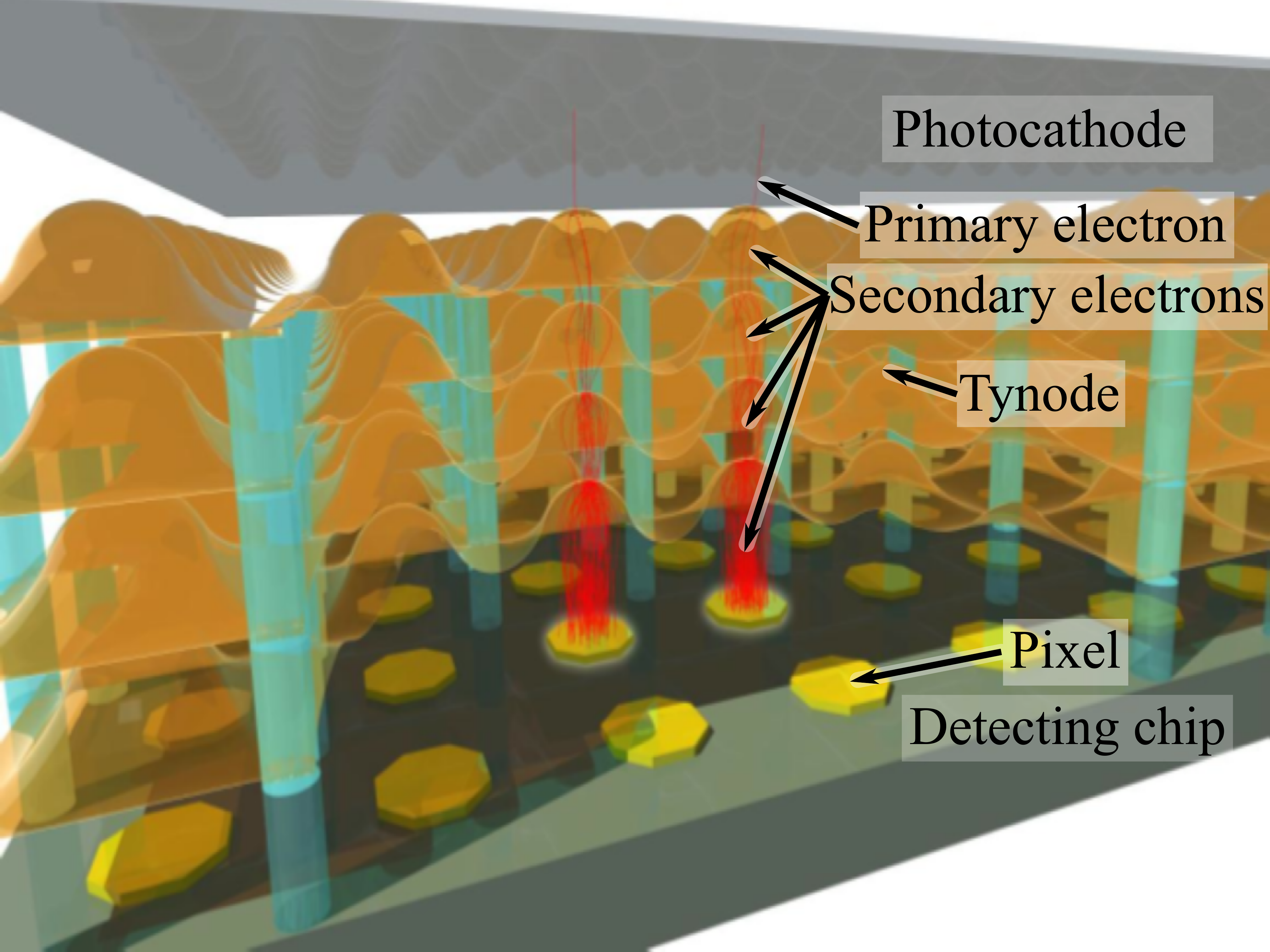}
	\caption{Artist impression of TiPC. A highly efficient photocathode emits a primary photoelectron after impact of a soft photon. This primary electron is multiplied by a stack of tynodes, which also focus the resulting secondary electrons onto the next tynode due to a dome-like shape. The resulting bunch of secondary electrons is detected by a fast CMOS chip. Figure adapted from \cite{Graafetal2013}.}\label{figTiPC-dream}
\end{figure}

The Timed Photon Counter (TiPC) is a concept photodetector with excellent spatio-temporal resolution and noise properties, potentially outperforming all photodetectors available to date \cite{Graafetal2013,Graafetal2017}. The envisioned device consists of three essential components as depicted in figure \ref{figTiPC-dream}. First, it utilises a highly efficient photocathode to generate a photoelectron upon impact of a soft photon ($0.1<\lambda<\SI{10}{\mu m}$). The photoelectron is multiplied by a stack of transmission dynodes, or tynodes, which are the second necessary component. These tynodes consist of a thin film with a thickness in the order of $\SI{10}{nm}$ in a dome-like shape. In these domes secondary electrons (SEs) are generated. Secondly, since we aim for a inter-tynode distance in the stack of $\sim\SI{100}{\mu m}$, the curvature of these domes causes the electrons to be focussed onto the next tynode, making TiPC suitable for use in magnetic fields. The stack of tynodes is positioned over a fast CMOS chip -- the third essential -- detecting the multiplied electron bunch with a predicted temporal resolution in the order of a picosecond and a spatial resolution that merely depends on the chip's pixel pitch and size.
\\

Previously, tynodes have been fabricated with a (transmission SE) yield up to $5.5$ \cite{Prodanovicetal2018,ProdanovicThesis,Chanetal,ChanetalII}. In this work we take the next step in the realisation of TiPC by combining two of the three main components. We characterise the yield of a tynode by use of a CMOS TimePix-chip. For now we still use an electron gun as a source of primary electrons (PEs).\\
In section \ref{secSetup} we describe the prototype TiPC used in this work in detail. The method for obtaining the yield of a tynode using our set-up is outlined in section \ref{secYieldDetermination}. By this method we obtain the yield of a tynode as a function of PE landing energy, as shown in section \ref{secResults}. In this section we also compare our results to results obtained by a Scanning-Electron-Microscope- (SEM-)based method described in \cite{ProdanovicThesis,Chanetal} using the same sample. Finally, some measurement features which are unrelated to the tynode's yield determination will be discussed in section \ref{secDiscussion}.

\section{Description of the set-up}\label{secSetup}
\begin{figure}[htbp]
	\centering
	\begin{subfigure}[b]{0.50\textwidth}
		\includegraphics[width=\textwidth]{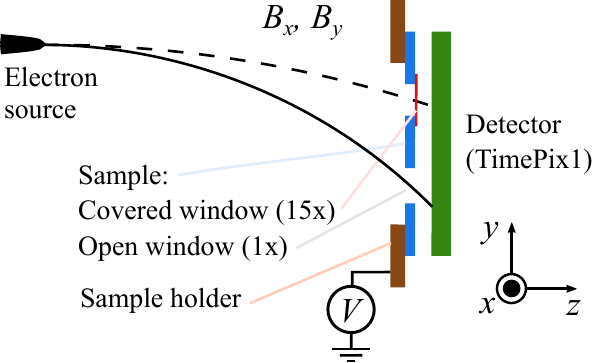}
		\caption{}\label{sfigSetup}
	\end{subfigure}\hspace{5mm}
	\begin{subfigure}[b]{0.45\textwidth}
		\includegraphics[width=\textwidth]{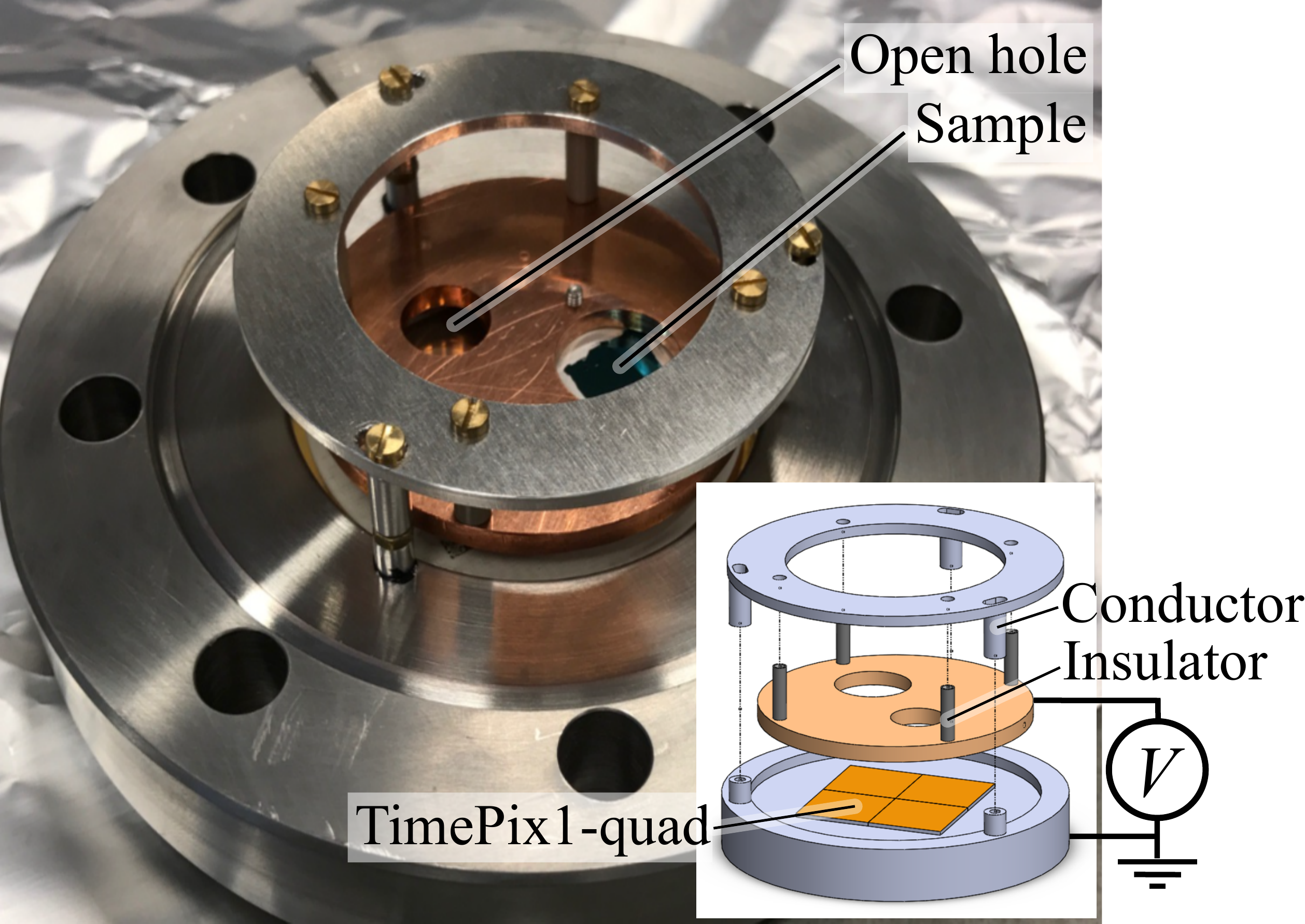}
		\caption{}\label{sfigSetupPhoto}
	\end{subfigure}\hspace{5mm}
	\caption{Overview of the set-up. (a) Schematic: An electron beam generated from a pulsed source is accelerated towards a sample. The sample, placed approximately $\SI{120}{mm}$ from the electron source, has $16$ windows of $1\times\SI{1}{mm^2}$, $15$ of which are spanned by a SE-generating thin film of tynodes (only $1$ shown explicitly in this figure) and one open window that serves as a reference. The beam can be scanned over the sample by an externally applied $\mbf{B}$-field generated using two sets of orthogonally positioned Helmholtz-coils. A quad TimePix$1$-detector is positioned $\SI{1}{mm}$ behind the sample to collect the electrons. The sample is placed in a sample holder that can be put at an elevated (negative) potential, such that the SEs leaving the thin films are accelerated towards the detector. (b) Detail of the sample holder and the quad TimePix$1$-detector. The Cu sample holder contains two holes. One is left open, whereas the sample is mounted in the other. The inset shows an exploded view.}\label{figSetup}
\end{figure}

The set-up consists of three vital parts: an electron source, a TimePix$1$-detector and a tynode sample. The set-up is schematically depicted in figure~\ref{sfigSetup}. Using a pulsed electron source\footnote{Kimball ELG-$2$ pulsed on the Wehnelt-cylinder with an in-house built pulse generator} we can generate a bunch of $0$ to $\SI{2}{keV}$-electrons. The pulse width can be varied between $\SI{20}{ns}$ and $\SI{100}{\mu s}$. The electron source contains an electrostatic lens for focussing the electron beam and two sets of electrostatic deflectors. It is noted that the deflectors should be used with caution due to severe aberration effects when large deflections are applied. To overcome these effects we added two pairs of Helmholtz-coils, which are mounted such that one controls the motion of the electrons in the (horizontal) $x$-direction, and the other controls the electrons' motion in the (vertical) $y$-direction. These coils deflect the electrons by their generated $\mbf{B}$-field during free flight over the approximately $\SI{120}{mm}$-distance between the electron source and the sample. The pairs of Helmholtz-coils are each controlled by a current source\footnote{Keithley Source-meter $2450$}, which is able to generate currents up/down to $\pm\SI{1}{A}$. Each of the coils has a diameter of $\SI{120}{mm}$ and carries $100$ turns of enamelled Cu wire. As such, we expect the magnitude of the generated $B_{x,y}$-fields to be approximately $\SI{1}{mT}$, which should be sufficient to scan the electron beam over the full sample.

A tynode sample is held by a Cu sample holder produced in-house, see figure \ref{sfigSetupPhoto}. Apart from the sample, this holder also contains a large open hole through which the beam of electrons hits the TimePix$1$-detector directly. This allows us to study the electron beam without the presence of a sample for e.g. focussing purposes. The sample holder is electrically isolated from the rest of the set-up, such that a non-zero potential can be applied to the sample only. By applying a negative potential using a voltage source$^{2}$, SEs generated in the sample are accelerated towards the detector. In this work we set the sample potential to $\SI{-200}{V}$. This, however, implies that the landing energy of the electrons is simultaneously reduced with respect to the energy by which the electrons are emitted from the source.\\
We detect electrons using a quad TimePix$1$-detector \cite{Llopartetal2007} in Time-over-Threshold (ToT-)mode. ToT is a (non-linear) measure for the charge received by the TimePix's pixels, and thus of the received amount of electrons. The $512\times512$ pixels form a square array with a pitch of $\SI{55}{\mu m}$. We set the TimePix$1$-detector to its $\SI{100}{MHz}$ clocking mode implying $1$ ToT-count corresponds to a ToT of $\SI{10}{ns}$.\\

\begin{figure}[htbp]
	\centering
	\begin{subfigure}[b]{0.3\textwidth}
		\includegraphics[width=\textwidth]{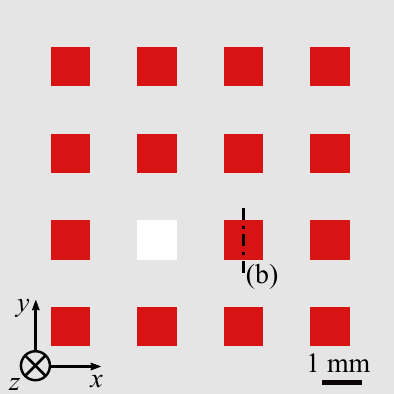}
		\caption{}\label{sfigSample}
	\end{subfigure}\hspace{8mm}
	\begin{subfigure}[b]{0.149\textwidth}
		\includegraphics[width=\textwidth]{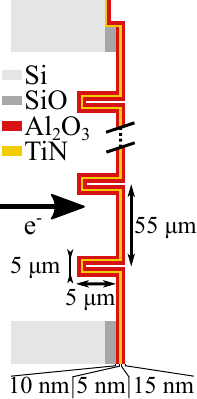}
		\caption{}\label{sfigMembraneSchematic}
	\end{subfigure}\hspace{6mm}
	\begin{subfigure}[b]{0.35\textwidth}
		\includegraphics[width=\textwidth]{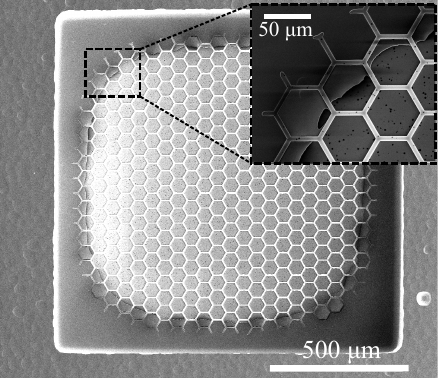}
		\caption{}\label{sfigMembraneSEM}
	\end{subfigure}
	\caption{Overview of the tynode sample. (a) Schematic overview of the full sample. It features $16$ windows of $1\times\SI{1}{mm^2}$. All windows except one are covered by a thin film featuring tynodes in which SEs are generated. The cross-section presented in (b) is indicated by a dash-dotted line. (b) Schematic overview of a film-covered window. The thin film consists of a TiN layer of $\SI{5}{nm}$-thickness sandwiched by two Al$_2$O$_3$ layers with a thickness of $10$ and $\SI{15}{nm}$ respectively. The Al$_2$O$_3$ layers serve to generate SEs, whereas the TiN layer provides some conductance in order to prevent charge-up of the thin film. In the results presented in this work, the electrons hit the film from the left as indicated. The sections of the film parallel to the direction of flight of the incoming electrons form the boundaries between separate cells of the tynodes and increase the structural integrity of the film. (c) SEM-image of a single film-covered window. The cells have the shape of hexagons and are ordered in a triangular lattice with a spacing of $\SI{55}{\mu m}$. The inset shows the hexagon cells in detail.}\label{figSample}
\end{figure}

The tynode sample consists of $16$ windows with approximate dimensions of $1\times\SI{1}{mm^2}$ positioned in a $4\times 4$-array, see figure \ref{sfigSample}. Of these $16$ windows, $15$ are covered by an Al$_2$O$_3$/TiN/Al$_2$O$_3$-trilayer thin film featuring the tynodes. The last window is left open as a reference. The covered windows are schematically depicted in figure~\ref{sfigMembraneSchematic} and their fabrication is described in \cite{ChanetalII}. SEs are created in the Al$_2$O$_3$ layers with respective thicknesses of $10$ and $\SI{15}{nm}$. The $\SI{5}{nm}$-thick TiN layer provides the necessary thin film conductance to prevent charge-up of the sample. With reference to figure \ref{sfigMembraneSchematic}, the horizontal sections of the films are the boundaries of separate tynode cells and provide structural support. A SEM-image of one of the covered windows is shown in figure~\ref{sfigMembraneSEM}. The vertical sections of the thin films are hexagonally shaped and are positioned on a triangular lattice with a lattice spacing of $\SI{55}{\mu m}$ -- the TimePix$1$'s pixel size.

The master instrument in our measurement system is an oscilloscope\footnote{Agilent DSO-X $3054$A}. Upon request this oscilloscope emits a trigger pulse to the pulse generator connected to the electron source. In turn, this pulse generator triggers a second pulse generator\footnote{Stanford DG$535$}, which opens and closes the shutter of the TimePix$1$-detector. The shutter is open for $\SI{118.10}{\mu s}$ corresponding to the maximum of $11810$ ToT-counts of the TimePix$1$-detector set to its $\SI{100}{MHz}$ clocking mode.

\section{Yield determination}\label{secYieldDetermination}
Because the TimePix$1$-detector is non-linear in its ToT-response \cite{Llopartetal2007,Akibaetal2012} and since our electron pulse spot size is larger than the $1\times\SI{1}{mm^2}$-windows we use a procedure of four steps to determine the yield of the thin films. To illustrate the method, we use an example measurement in which the electron beam emission energy equals $\SI{1.7}{keV}$. Since the sample is held at a potential of $\SI{-200}{V}$, the landing energy of the electrons is $\SI{1.5}{keV}$.\\
The four steps are the following: first we scan over the entire sample to find the positions of the $16$ windows on the TimePix$1$-detector. From the same measurement we obtain the approximate settings of the currents through the Helmholtz-coils necessary to hit the windows. In the second step we auto-focus the electron beam on each of the separate windows. We then perform a sweep of the electron pulse time to track the non-linear behaviour of the TimePix$1$-detector as a third step from which we determine the yield of the thin films in the final step. 

\subsection{Step $1$: $I_{x,y}$-scan}\label{ssecXYscan}
\begin{figure}[htbp]
	\centering
	\begin{subfigure}[b]{0.68\textwidth}
		\includegraphics[width=\textwidth,trim={0cm 1cm 0 -0.5cm},clip]{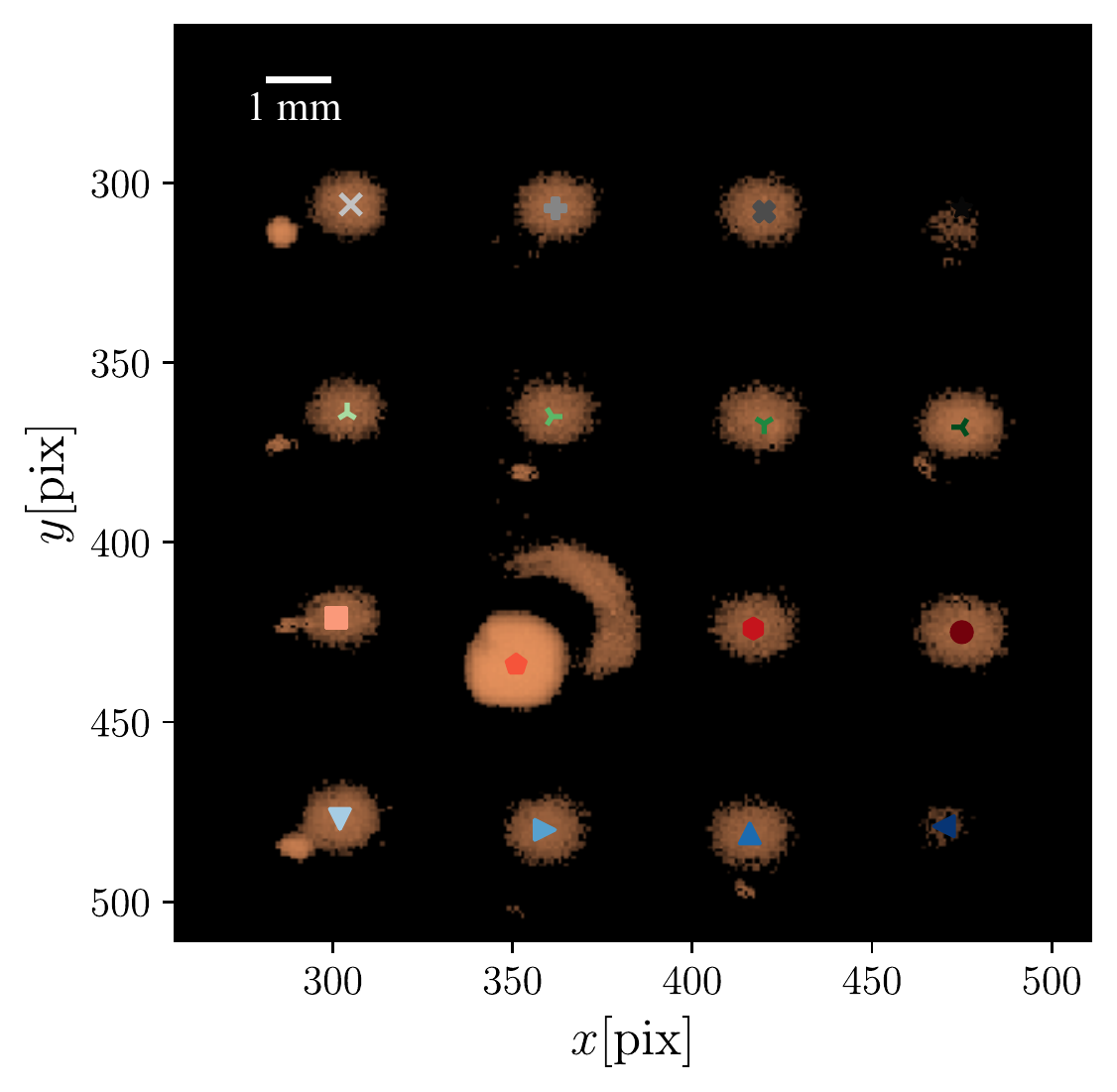}
		\caption{}\label{sfigSumFrame}
	\end{subfigure}\\
	\begin{subfigure}[b]{0.68\textwidth}
		\includegraphics[width=\textwidth,trim={0cm 1cm 0 -0.5cm},clip]{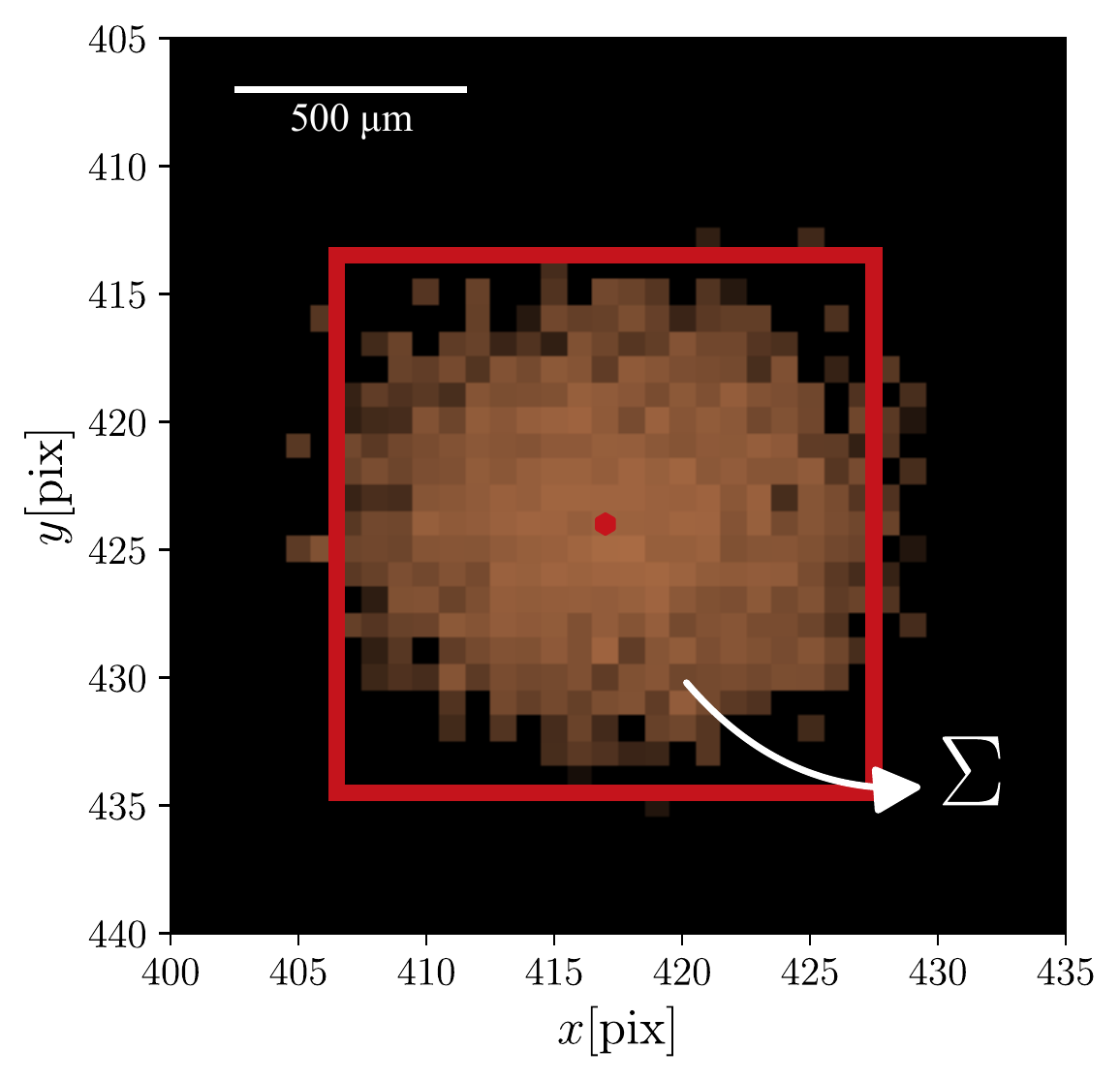}
		\caption{}\label{sfigSumFrame6}
	\end{subfigure}
	\caption{Typical sum-frames. (a) Sum-frame (log-scale) obtained by summing all measurement frames resulting from an $I_{x,y}$-scan. Three distinctive features are observed: a $4\times 4$ array of spots, satellites near some of the spots and a banana-shaped feature to the upper right of spot $(1,1)$ (see text for spot identifier nomenclature). Spot $(1,1)$ corresponds to the open window, whereas the other spots correspond to windows covered by a thin film. We identify the spot centres, indicated by dots, by fitting a two-dimensional Gaussian to the individual spots. (b) Zoom-in on spot $(1,2)$, which we will use as an example in this section. We define a square ROI centred on the spot centre with sides of $21$ pixels corresponding to the window size. As a measure for the charge reaching the detector for a certain frame, we use the number of summed ToT-counts within the ROI, indicated by $\Sigma$.}\label{figSumFrame}
\end{figure}

In the first step we scan the beam over the sample by adjusting the $I_{x,y}$-currents through the Helmholtz-coils generating the $B_{x,y}$-fields. For this measurement we use an approximate focus potential, in this case $\SI{660}{V}$. Setting the pulse time of the electrons to $\SI{1}{\mu s}$, we collect a measurement frame from the TimePix$1$-detector for each set of $I_{x,y}$. These measurement frames contain the number of ToT-counts per pixel collected during the time the TimePix$1$-shutter is open. A sum-frame as depicted in figure \ref{sfigSumFrame} is obtained by summing these measurement frames. The electron source is positioned approximately at the top right corner of the depicted sum-frame. This implies that the PEs hitting the sample have an in-plane velocity component directed to the bottom left.\\ 
In the sum-frame one may discern three distinctive features. The first feature is the array of $16$ spots corresponding to the $16$ windows. These spots will be designated by their row- (counting from below starting from $0$) and column-indices (counting from the left starting from $0$). Hence, the bright spot corresponding to the open window is spot $(1,1)$. Spots $(0,3)$ and $(3,3)$ are observed to be dimmer than the other spots, due to a partial overlap of the corresponding windows by the sample holder. The second set of features consists of the satellites appearing next to certain spots. These satellites have a preferential direction towards the bottom left with respect to the corresponding spots. We attribute these satellites to punch-through PEs or PEs passing through small cracks in the thin films as discussed in section \ref{secDiscussion}. Thirdly, we observe a banana-shaped feature at the top right of spot $(1,1)$. This feature may be attributed to SEs generated in the Si side walls of the sample, also discussed in section \ref{secDiscussion}.

\begin{figure}[htbp]
	\centering
	\begin{minipage}{0.50\textwidth}
		\includegraphics[width=0.99\textwidth,trim={2.5cm 0.5cm 0 0.5cm},clip]{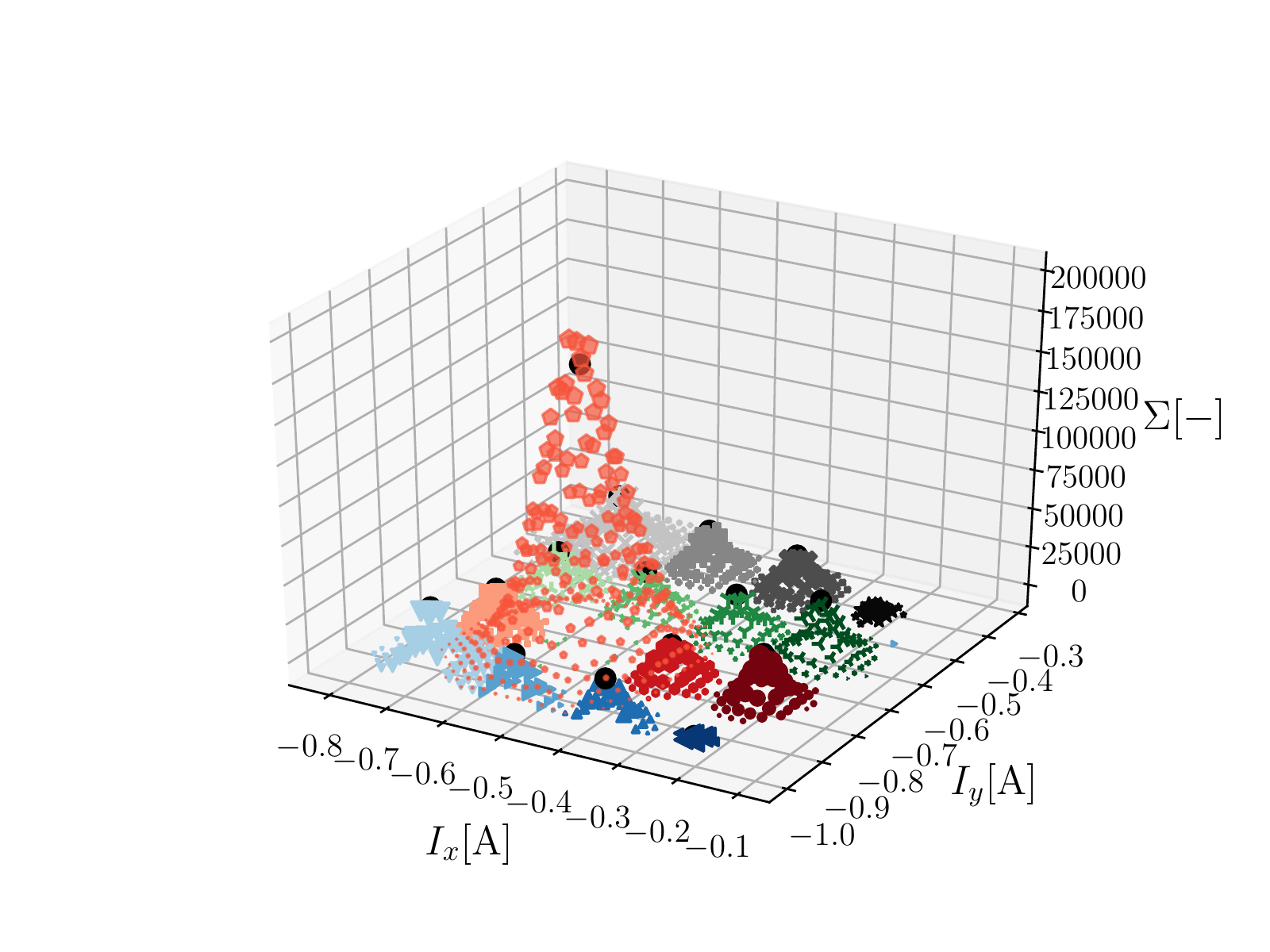}
		\caption{$\Sigma$ as a function of $I_{x,y}$ per spot. Since the electron beam footprint is larger than the area of the windows, we obtain results peaked at a certain $I_{x,y}$-setting. By fitting a two-dimensional Gaussian to the data, we obtain an approximation of the optimum $I_{x,y}$ for hitting a window, $I_{x,y}^{\text{opt*}}(i,j)$, indicated for each spot by a black dot. The colour code and markers in this figure are identical to the colour code chosen for the spot centres in figure \ref{sfigSumFrame}. The marker size within each peak is scaled for better data visualisation.}\label{figIxySigma}
	\end{minipage}\hfill
	\begin{minipage}{0.49\textwidth}
		\includegraphics[width=0.99\textwidth]{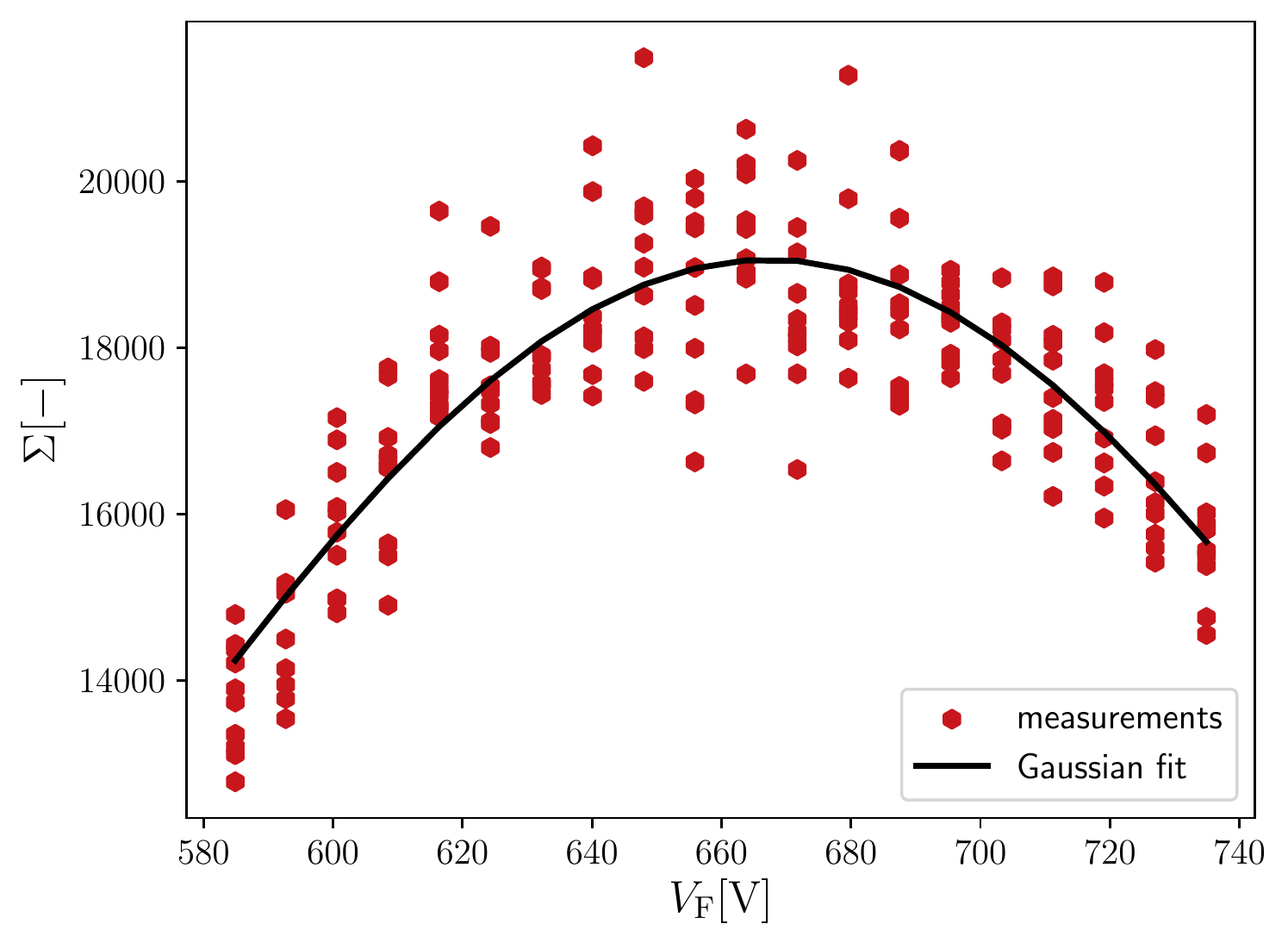}
		\caption{Auto-focus result: $\Sigma$ as a function of focus potential $V_{\text{F}}$ for spot $(1,2)$. Using $I_{x,y}^{\text{opt**}}(i,j)$ determined using a fine $I_{x,y}$-scan around the $I_{x,y}^{\text{opt*}}(i,j)$-values obtained in the $I_{x,y}$-scan of step $1$ (see figure \ref{figIxySigma}), we sweep the focus potential of the electron source in order to obtain the optimum focus for the windows, $V_{\text{F}}^{\text{opt}}(i,j)$, equalling $V_{\text{F}}$ for which $\Sigma$ is maximised. The dots correspond to the measured values and the line is a one-dimensional Gaussian fit to the data.}\label{figFscan}
	\end{minipage}
\end{figure}

From the sum-frame we determine the spot centres, as indicated in figure \ref{sfigSumFrame}, by fitting a two-dimensional Gaussian to the spots. We use the spot centres to define a square region of interest (ROI) around the spot centre corresponding to the square windows, see figure \ref{sfigSumFrame6}. The sides of the ROI are chosen to be $21$ pixels ($\approx\SI{1.2}{mm}$) in size, which is slightly larger than the window sides of $\SI{1}{mm}$. As such, we expect that all of the transmitted SEs from the windows with thin films and the PEs from the open window land inside the ROI. Contrarily, the contribution from the satellites, the banana-shaped feature and possible SEs generated by the PEs impacting the detector are outside of the ROI. We then calculate the sum of the ToT-counts within the relevant ROIs for each frame, $\Sigma$, in order to obtain the optimum $I_{x,y}$ for hitting the corresponding windows. $\Sigma$ is a measure for the total charge arriving in the TimePix$1$-detector per frame and per spot. An example of $\Sigma(I_x,I_y)$ is depicted in figure \ref{figIxySigma}. Fitting a two-dimensional Gaussian to the data, we obtain an estimate for the optimum $I_{x,y}$ per window, $I_{x,y}^{\text{opt*}}(i,j)$.

\subsection{Step $2$: Auto-focus}\label{ssecFF}
Using the ROIs and $I_{x,y}^{\text{opt*}}(i,j)$ from step $1$, we proceed by an auto-focus measurement to find the optimum focus potential of the electron source for each window. For this optimum, $\Sigma$ should be maximised. This measurement has three sub-steps, which are each typically performed at a pulse width of $\SI{1}{\mu s}$. First we perform a fine $I_{x,y}$-scan around $I_{x,y}^{\text{opt*}}(i,j)$ -- as described previously -- for a better approximation of this parameter, $I_{x,y}^{\text{opt**}}(i,j)$. Using the latter, we sweep the focus potential $V_{\text{F}}$ of the electron source and fit a one-dimensional Gaussian to the resulting $\Sigma$-values. For each $V_{\text{F}}$ we obtain $10$ measurement frames. This yields the optimum focus potential $V_{\text{F}}^{\text{opt}}(i,j)$, see figure \ref{figFscan}. Finally, a second $I_{x,y}$-scan is performed at $V_{\text{F}}^{\text{opt}}(i,j)$ in order to determine a final approximation of the optimum current settings, $I_{x,y}^{\text{opt}}(i,j)$.

\subsection{Step $3$: Pulse time variation}\label{ssecPT}
\begin{figure}[htbp]
	\centering
	\begin{minipage}[t]{0.47\textwidth}
		\includegraphics[width=0.97\textwidth]{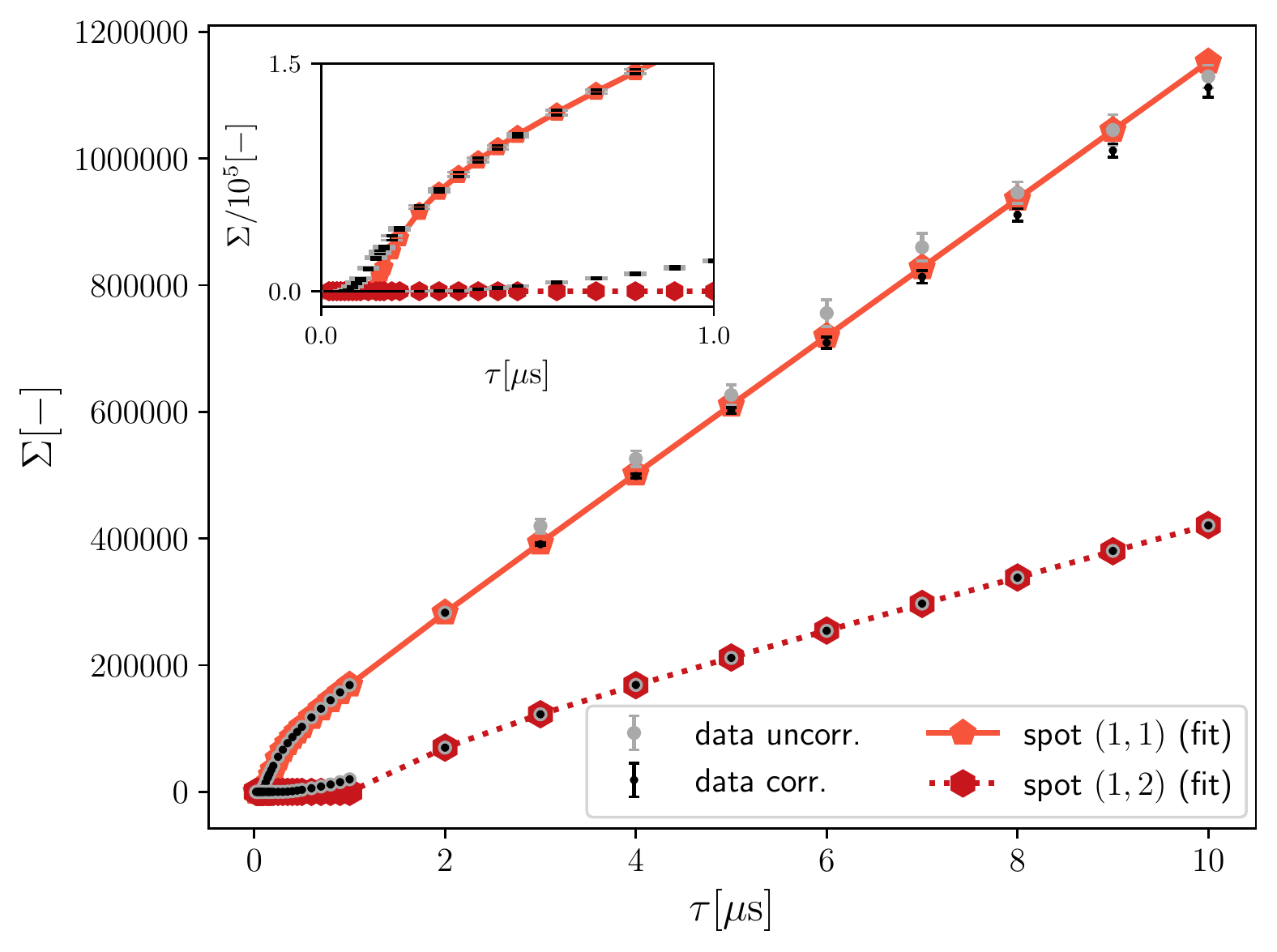}
		\caption{$\Sigma$ as a function of the pulse width $\tau$ for the open window (spot $(1,1)$) and one of the thin-film-covered windows (spot $(1,2)$). The data is fitted with equation~(\ref{eqSurrogateFunction}), leaving out the points where the TimePix$1$-detector shows onset or hot- and cold-pixel behaviour (see text). From division of the slopes of the linear part of the curves one may determine the yield of the thin film. The inset shows the $0$ to $\SI{1}{\mu s}$ domain of the data and the fitted curves in detail.}\label{figSvsTau}
	\end{minipage}
	\begin{minipage}[t]{0.47\textwidth}
		\includegraphics[width=0.97\textwidth]{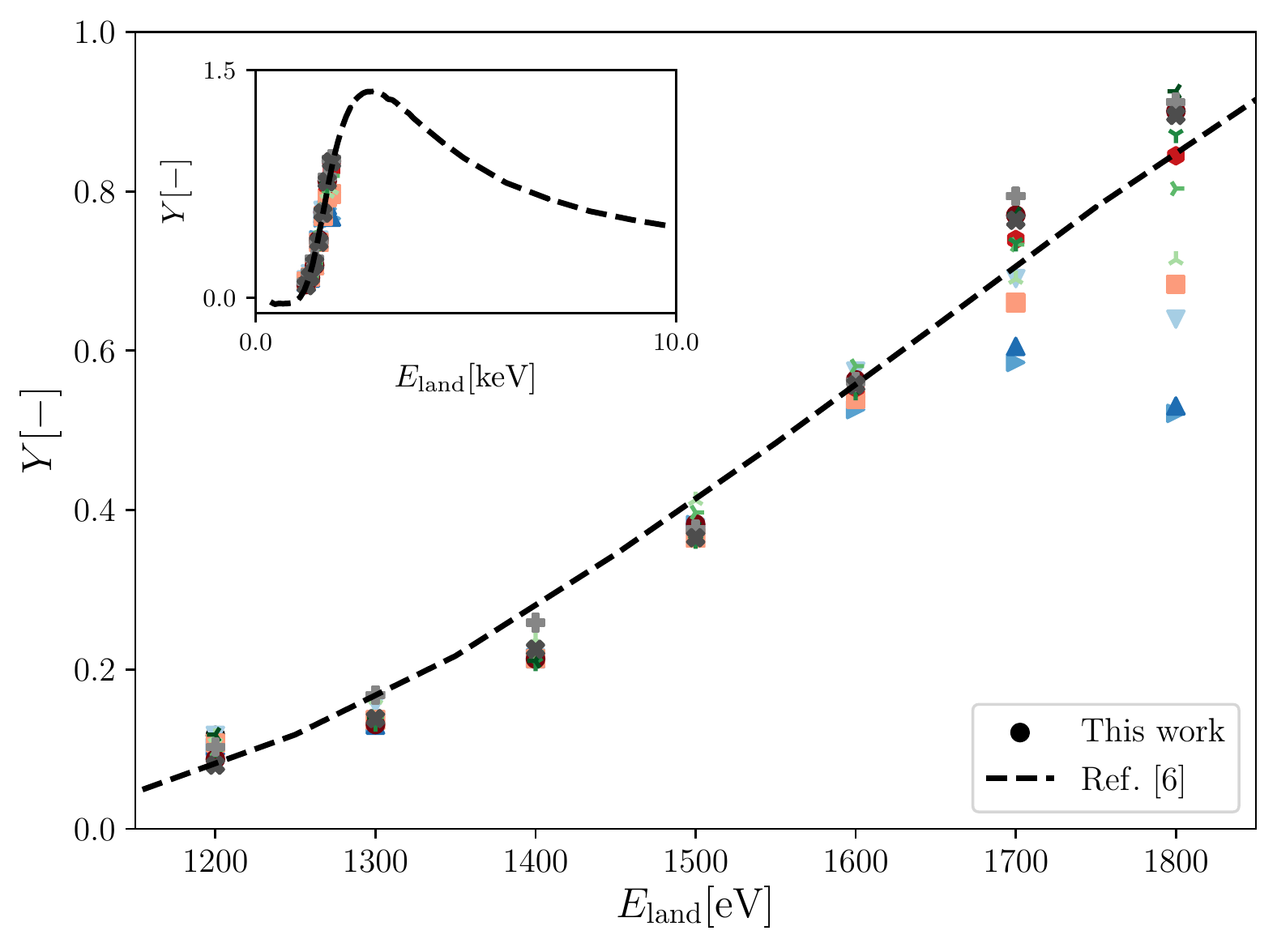}
		\caption{Yield of the thin films as function of the electron landing energy. The scattered dots correspond to our measured values, using the same colouring and markers as presented in figure \ref{figSumFrame} for spot identification. We compare our data to the yield obtained in \cite{ChanetalII} by a SEM-based method presented in \cite{ProdanovicThesis,Chanetal} (dashed line). The inset shows the full yield-curve obtained by this method.}\label{figY200}
	\end{minipage}
\end{figure}

As a final step, we set the electron beam focus to $V_{\text{F}}^{\text{opt}}(i,j)$ and $I_{x,y}$ to $I_{x,y}^{\text{opt}}(i,j)$ for each of the separate windows. We then perform a sweep over the pulse time of the electron pulses, $\tau$. For each $\tau$, $\Sigma$-values of $30$ measurement frames are obtained. In figure \ref{figSvsTau}, we have visualised the result for the open window, spot $(1,1)$, and for one of the windows covered by a thin film, spot $(1,2)$. As can be observed, the response of the TimePix$1$-detector is absent up to a certain $\tau$. In this regime the charge received by the pixels is below threshold. This regime is followed by a non-linear regime, in which the response rises non-linearly to a linear response. This behaviour is clearly visible in figure \ref{figSvsTau} for spot $(1,2)$ for $0.5<\tau<\SI{3.0}{\mu s}$ and for spot $(1,1)$ for $60<\tau<\SI{500}{ns}$. As can be seen the response for spot $(1,2)$ is linear for $\tau\geq\SI{3}{\mu s}$.\\ 
For spot $(1,1)$, however, the linearity of the response observed between $0.5\leq\tau\leq\SI{2}{\mu s}$ is broken for $\tau\geq\SI{3}{\mu s}$. In this regime the detector is found to be influenced by detector counting errors, which lead to ``hot'' and ``cold'' pixels. Hot (cold) pixels appear as random large ($0$) ToT-counts in random pixels for which a smaller (non-zero) response is expected based on the pixel's position within the spot. The effect of these pixels becomes visible in the uncorrected data presented for spot $(1,1)$ in figure \ref{figSvsTau}. The increased slope and uncertainty of the data points in the domain $3\leq\tau\leq\SI{8}{\mu s}$ is due to the influence of an excess of hot pixels. Later, for $\tau\geq\SI{8}{\mu s}$, the slope diminishes again due to an excess of cold pixels.\\
We can remove the hot-pixel behaviour by a frame post-processing method. We set a threshold value of $10\tau$ on the ToT of the pixels. This value was determined upon inspection. If the ToT of a pixel exceeds the threshold, we check whether the pixel's ToT-count is larger than twice the average ToT-count of the $8$ neighbouring pixels. If this is the case, we conclude the pixel is hot and set its ToT-count equal to the average Tot-count of the neighbouring pixels. The result of this correction is also depicted in figure \ref{figSvsTau}. Note that this method does not correct for the cold pixels described before.\\
The response can be described by the surrogate function \cite{Akibaetal2012}
\begin{equation}\label{eqSurrogateFunction}
	\Sigma(\tau)=\begin{cases}
		a_0\tau+a_1-\dfrac{a_2}{\tau-a_3}\quad&\text{if }\Sigma>0\\
		0&\text{else.}
	\end{cases}
\end{equation}
In this equation the parameters $a_i$ are experimentally determined fitting parameters. Originally this function is proposed for the description of the single-pixel response. However, neglecting an onset of the response makes it suitable as a description of multiple-pixel sum-response as well. One can observe this onset in figure \ref{figSvsTau} for spot $(1,1)$ in the domain $60\leq\tau\leq\SI{200}{ns}$ and for spot $(1,2)$ in the domain $0.5\leq\tau\leq\SI{1}{\mu s}$. The onset occurs due to the width of the electron pulse. For a Gaussian distributed pulse, the central pixels receive more electrons than the non-central ones. This implies that the central pixels may already be in their linear response regime, whereas the non-central ones are still in their non-linear or below-threshold regime. Both the onset as well as the hot- and cold-pixel effects have been left out in fitting the data to equation~(\ref{eqSurrogateFunction}) by choosing the appropriate $\tau$-domain.

\subsection{Step $4$: Yield}\label{ssecYield}
From the $\Sigma$-fits to the data presented in figure \ref{figSvsTau} we determine the yield of the windows covered with thin films. For a linear detector, the relation between $\Sigma$ and $\tau$ would be of the form
\begin{equation}\label{eqLinDet}
	\Sigma(\tau)=a_{\text{lin}}\tau.
\end{equation}
For large $\tau$ it is readily observed from equation~(\ref{eqSurrogateFunction}) that the TimePix$1$-detector behaves more and more like a linear detector. If one considers the limit $\tau\mapsto\infty$, it is found that equations~(\ref{eqSurrogateFunction}) and (\ref{eqLinDet}) are identical, since $a_0\tau\gg a_1$ in this limit (cf. figure \ref{figSvsTau}). This allows us to determine the yield of film $(i,j)$ as
\begin{equation}
	Y(i,j)=\dfrac{a_0(i,j)}{a_0(1,1)}.
\end{equation}

\section{Yield results}\label{secResults}
Using the approach outlined in the previous section, we measure the yield of the thin-film-covered windows as a function of landing energy. The results of this measurement are plotted in figure \ref{figY200} and compared to yield measurements obtained in \cite{ChanetalII} from a SEM-based method presented in \cite{ProdanovicThesis,Chanetal}. In this figure, we have removed the results for windows $(0,3)$, $(3,0)$ and $(3,3)$ (see figure \ref{sfigSumFrame}). Windows $(0,3)$ and $(3,3)$ are partly covered by the sample holder as mentioned, while film $(3,0)$ was found to yield measurement frames in which many of the pixels consistently showed hot-pixel behaviour. Due to this behaviour, $a_0(3,0)$ could not be determined properly. The reason for these consistent hot-pixels is unknown.\\
The agreement between the yield curve obtained in this work and that obtained in \cite{ChanetalII} is good. However, at landing energies of $1.7$ and $\SI{1.8}{keV}$ we observe some outliers. These correspond to spot row $0$ (blue triangles) and spot column $0$ (lightest). For row $0$ it was found that the generated $\mbf{B}$-field was insufficient to reach the maximum of the $\Sigma(I_x,I_y)$-peak (cf. figure \ref{figIxySigma}). We observed that this was not the case for the column-$0$ spots, but have not been able to identify another reason for these outlying results.\\
Disregarding the outliers at $E_{\text{land}}=1.7$ and $\SI{1.8}{keV}$, the main reason for the spread in the obtained yield results is due to inhomogeneous window dimensions. The etch rate of the deep-reactive-ion-etch procedure removing the Si in the windows varies with the position of the windows. This implies that the all windows have slightly different dimensions, which has not been taken into account in the analysis. 

\section{Discussion of other features}\label{secDiscussion}
\begin{figure}[htbp]
	\centering
	\includegraphics[width=0.50\textwidth]{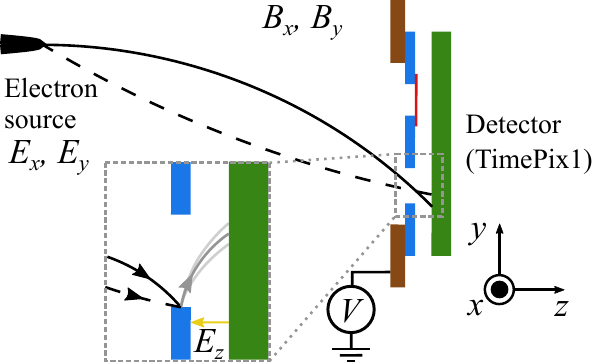}
	\caption{In order to interpret the satellites and the banana-shaped feature, we use the electrostatic deflectors of the electron source ($\mbf{E}$). The continuous and dashed lines indicate the trajectories of the electron beam for two different $(\mbf{B,E})$-settings, each passing through the open window. As such it can be understood that the location where PEs hit the TimePix$1$-detector are influenced by the trajectory's $(\mbf{B,E})$-setting. The inset shows the corresponding trajectories of the PEs hitting the sample wall, which we interpret to be responsible for the banana-shaped feature. The emitted SEs are accelerated towards the detector by the $E_z$-field resulting from the voltage applied onto the sample holder. Since the energy of these SEs is small and the sample is close ($\sim\!\!\!\SI{1}{mm}$) to the detector, they are hardly influenced by the $\mbf{B}$-field. Hence, it is understood that the location where SEs hit the detector is not influenced by the $(\mbf{B,E})$-settings.}\label{figSetupBE}
\end{figure}

As mentioned, the sum-frame depicted in figure \ref{sfigSumFrame} contains several other features apart from the $16$ spots. We discuss the nature of these features here. As shown schematically in figure \ref{figSetupBE} we expect two types of features: those connected to PEs, which are expected to move if the (solid) angle of incidence of the incoming electrons is altered and those connected to SEs, for which this is not the case. The reason for this behaviour is the fact that the SEs have a low energy and are formed at approximately $\SI{1}{mm}$ from the TimePix$1$-detector. Therefore their trajectories are hardly influenced by the $\mbf{B}$-field, but only by the $E_z$-field due to the sample potential. Hence, changing the angle of incidence of the PEs by using the electrostatic deflectors of the electron source should reveal whether features are due to PEs or SEs.\\

\begin{figure}[htbp]
\centering
	\begin{subfigure}[b]{0.49\textwidth}
	\includegraphics[width=0.89\textwidth]{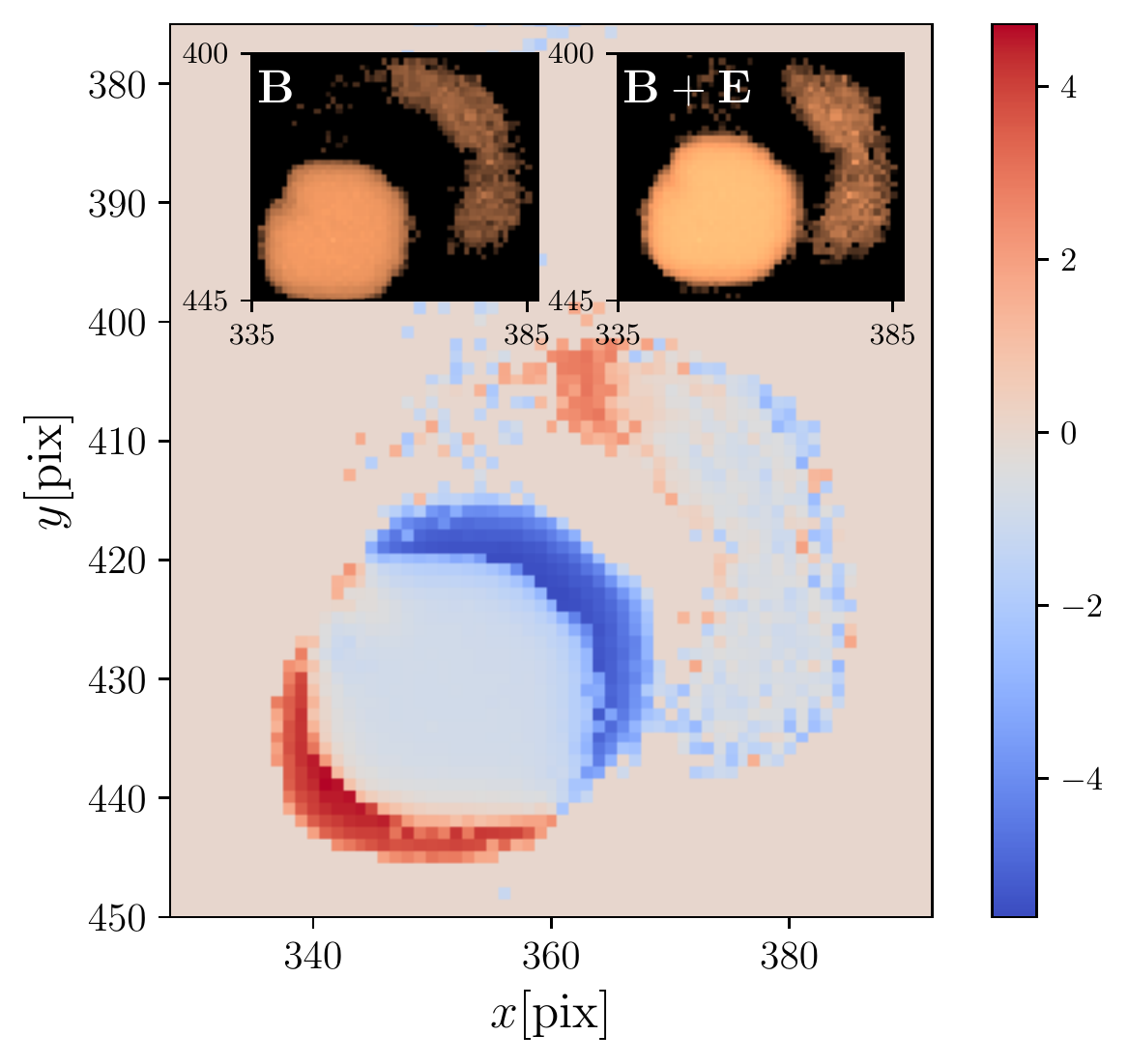}
	\caption{}\label{sfigBanana}
	\end{subfigure}
	\begin{subfigure}[b]{0.43\textwidth}
	\includegraphics[width=\textwidth]{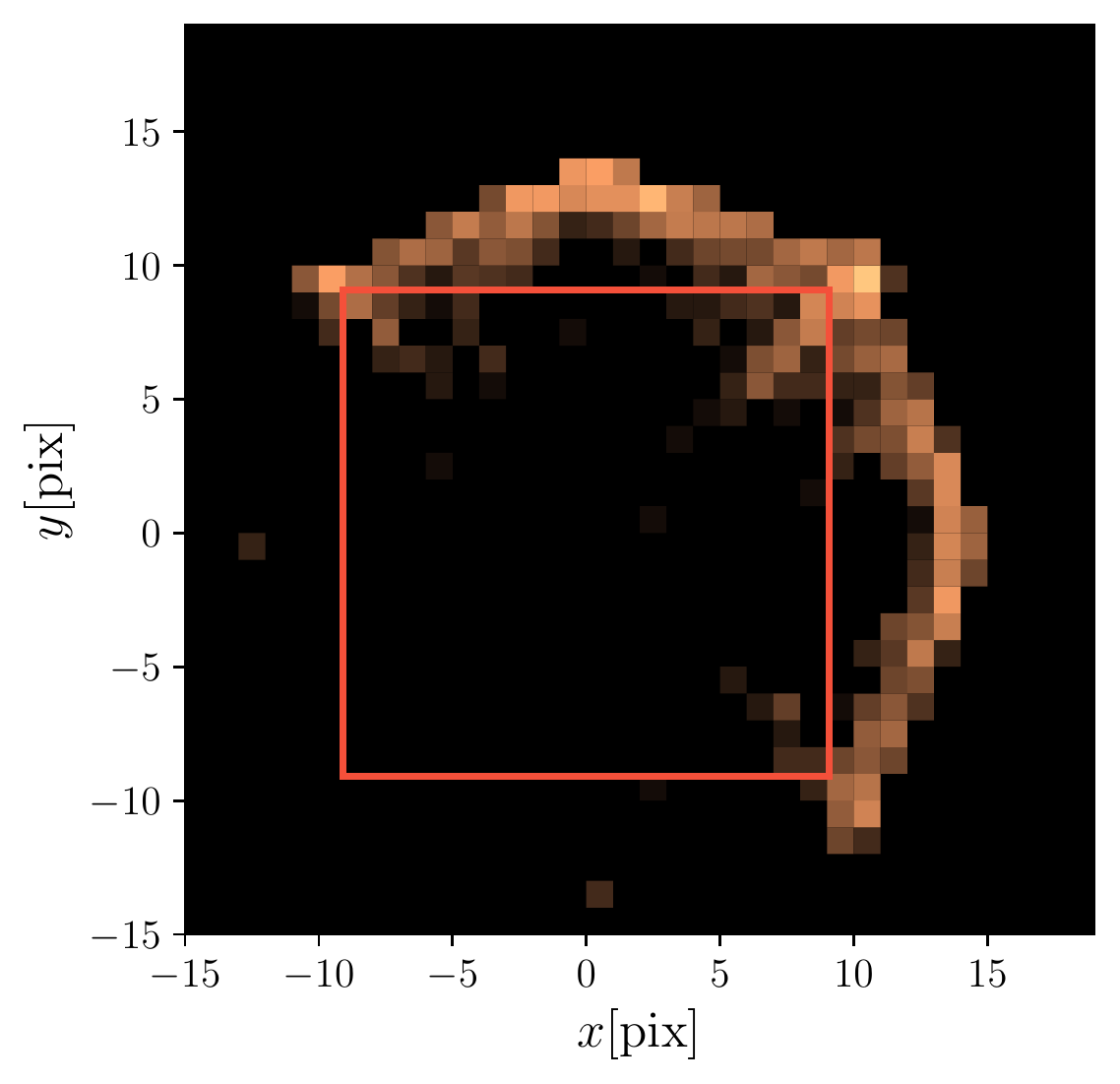}
	\caption{}\label{sfigBananaSim}
	\end{subfigure}
	\caption{Interpretation of the banana-shaped feature in figure \ref{sfigSumFrame}. (a) Differential sum-frame for spot $(1,1)$. The insets show the respective sum-frames (log-scale) from an $I_{x,y}$-scan without using the $x,y$-deflectors of the electron source (left) and from a scan using the $x,y$-deflectors in which we fixed $(I_x,I_y)=(-0.14,-0.23)\SI{}{A}$ (right). The main figure shows the difference of the insets. We observe the main spot to shift with the $(\mbf{B},\mbf{E})$-settings, since this feature results from PEs. The banana-shaped feature does not shift, from which we conclude it is due to SEs. (b) Monte-Carlo simulation result for the banana-shaped feature. The approximate position of the Si walls is indicated using the red square. We assume electrons with an initial energy of $0$ to $\SI{0.1}{eV}$ (uniform distribution) to depart at a random position and with a random orientation from the walls of the window irradiated with PEs (i.e., the left and lower wall). The $\sim 10^5$ simulated trajectories show a banana-shaped feature in accordance with (a). Electrons departing close to the edges of the window have been omitted, because of edge effects in the simulated electric field. Simulated pixels in which only a small amount of electrons land have been neglected in accordance with the non-linear behaviour of the TimePix$1$-detector.}\label{figBanana}	
\end{figure}

We have performed this measurement for the open window (spot $(1,1)$) and a covered window featuring a satellite (spot $(3,0)$; it should be noted that in this experiment -- performed in a previous experimental run -- the detector did not show the hot-pixel behaviour mentioned in section \ref{secResults}.). As mentioned in section \ref{secYieldDetermination} and shown schematically in the inset of figure \ref{figSetupBE}, we attribute the banana-shaped feature to SEs generated in the Si walls of the open window upon impact of the PEs. As can be observed in figure \ref{sfigBanana}, the banana-shaped feature is found not to shift upon changing the PEs angle of incidence, pointing to an SE-nature of these electrons. To further verify the interpretation, we performed a Monte-Carlo simulation in \textsc{Python} of the trajectories of low-energy SEs leaving the wall of the window within the electro-magnetic environment featured in the experiment. The electric field was estimated using a finite difference method and the (classical) electron trajectories were calculated using a standard fourth-order Runge-Kutta method. The result of the Monte-Carlo simulation is shown in figure \ref{sfigBananaSim}. From the similarity in the measured sum-frame and the simulations, we conclude that the banana-shaped feature can be attributed to SEs from the side walls of the open window.\\

\begin{figure}[htbp]
	\centering
	\includegraphics[width=0.49\textwidth]{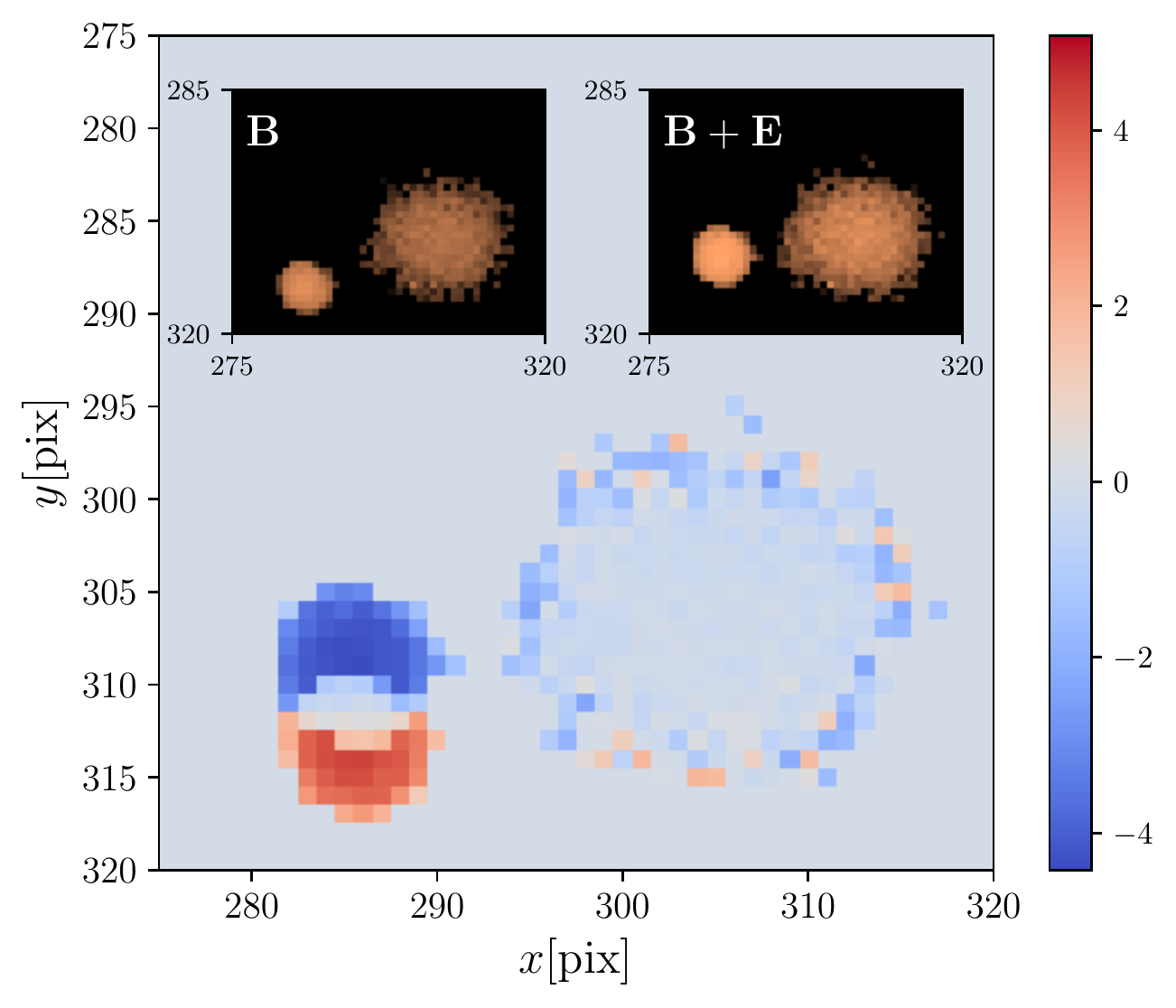}
	\caption{Interpretation of the satellites in figure \ref{sfigSumFrame}. Differential sum-frame for spot $(3,0)$. The insets show the respective sum-frames (log-scale) of an $I_{x,y}$-scan for which the $y$-deflector potential is set to $\SI{0}{V}$ (left) and $\SI{150}{V}$ (right). The main figure shows the difference of the insets. It is clear that the satellite shifts in positive $y$-direction due to the positive $y$-deflection, implying the satellite results from PEs. The spot is observed not to shift, indicating it results from SEs.}\label{figSatellite}
\end{figure}

In figure \ref{figSatellite} we show the result of a similar experiment performed on spot $(3,0)$. As is directly observed, the main SE-spot does not shift with the angle of incidence of the PEs, whereas the satellite does shift. This indicates that the satellites are due to PEs. In figure \ref{sfigSumFrame} it is observed that the satellites do not have a fixed position with respect to the main spots, and are absent for some spots. This implies their generation is due to a process which is random to a certain extent. The satellites could form e.g. due to micro-cracks in the vulnerable thin films through which PEs pass, or due to pinch-through PEs that experience slight inhomogeneities in the films due to non-perfect fabrication.

\section{Conclusions and Outlook}
At this stage of the development of the Timed Photon Counter (TiPC), we have combined a pulsed electron source, a tynode and a TimePix1-chip to build a TiPC prototype. Using this set-up we have determined the transmission secondary electron yield of the tynode as a function of electron landing energy. We compared our data with measurements obtained previously by a scanning-electron-microscope method and find a good agreement between both approaches.\\
In essence, our set-up is able to measure the effect of an extracting field on the secondary electron emission of the tynodes. In order to perform such a study, the sample voltage source should be replaced by a model which can deliver higher voltages than our current voltage source.\\
As a next step, we will update the TimePix$1$-detector to a TimePix$3$-detector. This detector can be calibrated more easily for absolute charge, making yield measurements more straightforward. Secondly, since the time resolution of the TimePix$3$-detector is better than that of the current TimePix$1$-detector ($\SI{1.56}{ns}$ vs. $\SI{25}{ns}$), we might be able to perform the timing measurements that are at the core of TiPC as a means to distinguish primary from secondary electrons.\\ 
Additionally, we will measure different high-yield samples. Simultaneously we shall develop a method to align multiple tynodes in a stack, such that their combined yield is sufficient to detect the arrival of a single primary electron. Making the tynode films thinner, the yield maximises at lower primary electron energy. Such tynodes are more suitable for TiPC, as the inter-tynode voltage may be reduced.\\
Apart from these developments we will work on a highly efficient photocathode and by combining this photocathode, the stack of tynode films and the TimePix$3$-detector, TiPC may become a reality in the near future.

\subsection*{Acknowledgements}
We would like to thank P. Timmer, G.W. Gotink, H. Verkooijen, J.H.M. van der Linden and J. van der Cingel for technical support. We are grateful to P.M. Sarro and V. Prodanovi\'c for their general contribution to the development of the tynode. This research was funded by ERC-Advanced $2012$/$320764$/MEMBrane, and by 
ERC-ATTRACT $2019$ CERN/HighQE.\\

\bibliographystyle{unsrt}
\bibliography{TyTest_JINST_arxiv.bbl} 

\begin{thebibliography}{1}

\bibitem{Graafetal2013}
H.~van~der Graaf, M.~A. Bakker, H.~W. Chan, E.~Charbon, F.~Santagata, P.~M.
  Sarro, and D.~R. Schaart.
\newblock The tipsy single soft photon detector and the trixy ultrafast
  tracking detector.
\newblock {\em J. Instrum.}, 8:C01036, 2013.

\bibitem{Graafetal2017}
H.~{van der Graaf}, H.~Akhtar, N.~Budko, H.~W. Chan, C.~W. Hagen, C.~C.~T.
  Hansson, G.~N\"utzel, S.~D. Pinto, V.~Prodanovi\'c, B.~Raftari, P.~M. Sarro,
  J.~Sinsheimer, J.~Smedley, S.~Tao, A.~M. M.~G. Theulings, and K.~Vuik.
\newblock The tynode: A new vacuum electron multiplier.
\newblock {\em Nucl. Instrum. Meth. A}, 847:148--161, 2017.

\bibitem{Prodanovicetal2018}
V.~Prodanovi{\'{c}}, H.~W. Chan, H.~{van der Graaf}, and P.~M. Sarro.
\newblock Ultra-thin alumina and silicon nitride {MEMS} fabricated membranes
  for the electron multiplication.
\newblock {\em Nanotechnology}, 29:155703, 2018.

\bibitem{ProdanovicThesis}
V.~Prodanovi{\'{c}}.
\newblock {\em Ultra-thin mems fabricated tynodes for electron multiplication}.
\newblock PhD thesis, Delft University of Technology, 2019.

\bibitem{Chanetal}
H.~W. Chan, V.~Prodanovi{\'{c}}, A.~M. M.~G. Theulings, C.~W. Hagen, P.~M.
  Sarro, and H.~{van der Graaf}.
\newblock Secondary electron emission from multi-layered
  {T}i{N}/{A}l$_2${O}$_3$ transmission dynodes.
\newblock \emph{e-print} arXiv:$2008.08997$, $2020$.

\bibitem{ChanetalII}
H.~W. Chan, V.~Prodanovi{\'{c}}, T.~ten Bruggencate, C.~W. Hagen, P.~M. Sarro,
  and H.~{van der Graaf}.
\newblock Mechanical meta-material of multi-layered
  {A}l$_2${O}$_3$/{T}i{N}/{A}l$_2${O}$_3$ film as large-surface transmission
  dynode.
\newblock \emph{e-print} arXiv:$2008.09054$, $2020$.

\bibitem{Llopartetal2007}
X.~Llopart, R.~Ballabriga, M.~Campbell, L.~Tlustos, and W.~Wong.
\newblock {Timepix, a $65$k programmable pixel readout chip for arrival time,
  energy and/or photon counting measurements}.
\newblock {\em Nucl. Instrum. Meth. A}, 581:485--494, 2007.
\newblock [Erratum: Nucl.Instrum.Meth.A 585, 106--108 (2008)].

\bibitem{Akibaetal2012}
K.~Akiba, M.~Artuso, R.~Badman, A.~Borgia, R.~Bates, F.~Bayer, M.~{van
  Beuzekom}, J.~Buytaert, E.~Cabruja, M.~Campbell, P.~Collins, M.~Crossley,
  R.~Dumps, L.~Eklund, D.~Esperante, C.~Fleta, A.~Gallas, M.~Gandelman,
  J.~Garofoli, M.~Gersabeck, V.~V. Gligorov, H.~Gordon, E.~H.~M. Heijne,
  V.~Heijne, D.~Hynds, M.~John, A.~Leflat, L.~{Ferre Llin}, X.~Llopart,
  M.~Lozano, D.~Maneuski, T.~Michel, M.~Nicol, M.~Needham, C.~Parkes,
  G.~Pellegrini, R.~Plackett, T.~Poikela, E.~Rodrigues, G.~Stewart, J.~Wang,
  and Z.~Xing.
\newblock Charged particle tracking with the timepix {ASIC}.
\newblock {\em Nuc. Instrum. Meth. A}, 661:31--49, 2012.

\end{thebibliography}
\end{document}